\RequirePackage{ifpdf}
\ifpdf 
\documentclass[pdftex]{sigma}
\else
\documentclass{sigma}
\fi

\begin{document}
\allowdisplaybreaks

\renewcommand{\PaperNumber}{007}

\FirstPageHeading

\ShortArticleName{On the GUEs with Additional Symmetries}

\ArticleName{On the Gaussian Random Matrix Ensembles\\ with
Additional Symmetry Conditions}

\Author{Vladimir VASILCHUK} \AuthorNameForHeading{V.~Vasilchuk}

\Address{B. Verkin Institute for Low Temperature Physics and
Engineering,\\ 47 Lenin Ave.,
 Kharkiv,  61103 Ukraine}

\Email{\href{mailto:vasilchuk@ilt.kharkov.ua}{vasilchuk@ilt.kharkov.ua}}

\ArticleDates{Received October 31, 2005, in final form January 06,
2006; Published online January 21, 2006}

\Abstract{The Gaussian unitary random matrix ensembles satisfying
some additional symmetry conditions are considered. The effect of
these conditions on the limiting normalized counting measures and
correlation functions is studied.}


\Classification{15A52; 60A10; 82B41}

\section{Introduction and main results}

Let us consider a standard $2n\times 2n$ Gaussian Unitary Ensemble
(GUE) of Hermitian random matrices $W_{n}$:
\begin{gather}
W_{n}=W_{n}^{\dagger },\qquad \left( W_{n}\right)
_{xy}=\frac{1}{\sqrt{2n}}\left( \xi _{xy}+i\eta _{xy}\right) ,
\label{GUE}
\end{gather}
where $\xi _{xy}$, $\eta _{xy}$, $x,y=-n,\ldots ,-1,1,\ldots ,n$
are i.i.d.\ Gaussian random variables with zero mean and variance
$1/2$. Consider also the \textit{normalized eigenvalue counting
measure} (NCM) $N_{n}$ of the ensemble (\ref{GUE}), defined for
any Borel set $\Delta \subset \mathbb{R}$ by the formula
\begin{gather}
N_{n}(\Delta )=\frac{\#\{\lambda _{i}\in \Delta \}}{2n},
\label{NCM}
\end{gather}
where $\lambda _{i}$, $i=1,\ldots ,2n$ are the eigenvalues of
$W_{n}$.

Suppose now that ensemble (\ref{GUE}) has also an additional
symmetry of negative (positive) indices $x$ and $y$. We consider
four different cases of symmetry:
\begin{gather}
1. \quad \left( W_{n}\right) _{xy}=\left( W_{n}\right) _{-y-x},
\label{Bel}
\\
2. \quad \left( W_{n}\right) _{xy}=\left( W_{n}\right) _{-x-y},
\label{Pas}
\\
3. \quad \left( W_{n}\right) _{xy}=\left( W_{n}\right) _{-xy},
\label{Me}
\\
4. \quad \left( W_{n}\right) _{xy}=\left( W_{n}\right) _{y-x}.
\label{Em}
\end{gather}

The Gaussian unitary ensemble and Gaussian orthogonal ensemble
(GOE) was considered in numerous papers (see e.g.~\cite{KKP:96}).
The Gaussian unitary ensemble with additional symmetry of
type~(\ref{Bel}) was proposed in the papers~\cite{BMR:03, DR:03}
as an approach to the weak disorder regime in the Anderson model.
This ensemble was also considered in the
papers~\cite{D:04,SSB:05}. In all these papers ensemble
(\ref{Bel}) was called as flip matrix model and studied by some
supersymmetry approach and moments method. In this paper an
approach is proposed that is simpler and the same for all four
cases (\ref{Bel})--(\ref{Em}). This approach is a version of
technique initially proposed in \cite{Ma-Pa:67} and developed in
the papers \cite{KKPS:92,KKP:96,Kh:96,KhPaVa:95}.

Using this technique we obtain the following results.

First two ensembles (\ref{Bel}) and (\ref{Pas}) are GOE-like.

\begin{proposition}
The NCMs $N_{n}^{(1)}$ and $N_{n}^{(2)}$ of the ensembles
\eqref{Bel} and \eqref{Pas} converge weakly with probability $1$
to the semi-circle law $N_{sc}$
\[
N_{sc}(\mathrm{d}\lambda )= (2\pi)^{-1}\sqrt{4-\lambda^2}\chi
_{\left[ -2,2\right] }(\lambda )\mathrm{d}\lambda
\]
and the $n^{-1}$-asymptotics of the correlation functions
\[
F_{n}^{(i)}(z_{1},z_{2})=\mathbb{E}\left\{ \left(
g_{n}^{(i)}(z_{1})- \mathbb{E}g_{n}^{(i)}(z_{1})\right) \left(
g_{n}^{(i)}(z_{2})-\mathbb{E} g_{n}^{(i)}(z_{2})\right) \right\}
,\qquad i=1,2
\]
of their Stieltjes transforms
\[
g_{n}^{(i)}(z)=\int_{-\infty }^{\infty }\frac{N_{n}^{(i)}(\mathrm{d}\lambda )%
}{\lambda -z},\qquad \mathrm{Im}\, z>0,\quad i=1,2
\]
coincide with corresponding $2n\times 2n$-GOE asymptotic $\left(
2n\right) ^{-2}S(z_{1},z_{2})$~{\rm \cite{KKP:96}}:
\begin{gather}
F_{n}^{(i)}(z_{1},z_{2})=\left( 2n\right)
^{-2}S(z_{1},z_{2})+o\left(n^{-2}\right),
\nonumber\\
S(z_{1},z_{2})=\frac{2}{\left( 1-f_{sc}^{2}(z_{1})\right) \left(
1-f_{sc}^{2}(z_{2})\right) }\left(
\frac{f_{sc}(z_{1})-f_{sc}(z_{2})}{z_{1}-z_{2}}\right) ^{2},
\label{CGOE}
\end{gather}
where
\[
f_{sc}(z)=\int_{-\infty }^{\infty }\frac{N_{sc}(\mathrm{d}\lambda
)}{\lambda -z},\qquad \mathrm{Im}\, z>0
\]
is the Stieltjes transform of the semi-circle law $N_{sc}$.
\end{proposition}

\bigskip

The fourth ensemble (\ref{Em}) is GUE-like:\bigskip

\begin{proposition}
The NCM $N_{n}^{(4)}$ of the ensemble \eqref{Em} converges weakly
with probability $1$ to the semi-circle law $N_{sc}$ and the
$n^{-1}$-asymptotic of the correlation function of its Stieltjes
transform coincides with \eqref{CGOE}
 divided by $2$ (i.e.\ GUE asymptotic).
\end{proposition}

As for the third ensemble, the additional symmetry produces new
limiting NCM and correlation function:

\begin{theorem}
The NCM $N_{n}^{(3)}$ of the ensemble \eqref{Me} converges weakly
with probability $1$ to the limiting non-random measure $N$
\begin{gather}
N(d\lambda )=\frac{1}{4}\delta (\lambda )d\lambda +\frac{1}{4\pi
}\sqrt{ 6-(\lambda ^{2}-\lambda ^{-2})}\chi _{\left[ -\lambda
_{+},-\lambda _{-} \right] \cup \left[ \lambda _{-},\lambda
_{+}\right] }(\lambda )d\lambda , \label{MeN}
\end{gather}
where $\lambda _{\pm }=\sqrt{3\pm 2\sqrt{2}}$ and the
$n^{-1}$-asymptotic of the correlation function of its Stieltjes
transform is given by the formula
\begin{gather}
F_{n}^{(3)}(z_{1},z_{2})=\left( 2n\right)
^{-2}C(z_{1},z_{2})+o\left(n^{-2}\right),
\nonumber\\
C(z_{1},z_{2})=\left(2\frac{f^2(z_1)+f^2(z_2)}{f(z_1)f(z_2)(z_1-z_2)^2}+
\frac{z_2f(z_2)+z_1f(z_1)}{2z^{2}_1z^{2}_2f(z_1)f(z_2)} \right)\!\!\prod\limits_{k=1,2}\left( z_{k}+z_{k}^{-1}+4f(z_{k})%
\right)^{-1}  , \label{MeC}
\end{gather}
where $f(z)$ is Stieltjes transform of the limiting measure $N$.
\end{theorem}

This result is somewhat unexpected for the Hermitian Gaussian
random matrix ensemble with the rather large number (of the order
$n^{2}$) of independent random parameters. But it shows how much
the additional symmetry may affect the asymptotic behavior of the
eigenvalues.

\section{The limiting NCMs}
In this section we consider the limiting normalized countable
measures of the ensembles (\ref{Bel})--(\ref{Em}).

In that follows we use the notations
\begin{gather*}
G(z)=\left( W_{n}-z\right) ^{-1},\qquad
\hat{g}(z)=\frac{1}{2n}\sum\limits_{j=-n}^{n}G_{j-j}(z),
\\
g(z)=\frac{1}{2n}\mathrm{Tr}\,
G(z)=\sum\limits_{j=-n}^{n}G_{jj}(z),
\end{gather*}
and $\langle \cdot \rangle $ to denote the average over GUE. We
also use the resolvent identity
\[
G(z)=-z^{-1}I+z^{-1}W_{n}G(z)
\]
and the Novikov--Furutsu formula for the complex Gaussian random
variable $\zeta =\xi+i\eta$
with zero mean and variance 1, and for the continuously differentiable function $%
q(x,\overline{x}) $%
\begin{gather}
\mathbb{E}\zeta q(\zeta,
\overline{\zeta})=\mathbb{E}\frac{\partial}{\partial\overline{\zeta}}q(\zeta
,\overline{\zeta}), \label{NF}
\end{gather}
where $\frac{\partial}{\partial\overline{\zeta}}=\frac{1}{2}
\left(\frac{\partial}{\partial\xi}+i\frac{\partial}{\partial\eta}\right)$.

We will perform our calculations in parallel for all four
ensembles. First, let us observe that properties
(\ref{Bel})--(\ref{Me}) are valid not only for the matrices of
ensembles (\ref{Bel})--(\ref{Me}) but for their powers and 
hence also for their resolvents. Indeed, using induction by $m$
and the symmetry of summing index we obtain:
\begin{gather*}
1. \quad \left( W_{n}^{m+1}\right) _{jk} =\sum_{l=-n}^{n}\left(
W_{n}\right) _{jl}\left( W_{n}^{m+1}\right)
_{lk}=\sum_{l=-n}^{n}\left( W_{n}\right)
_{j-l}\left( W_{n}^{m+1}\right) _{-lk} \\
\phantom{1. \quad \left( W_{n}^{m+1}\right) _{jk}}{}
=\sum_{l=-n}^{n}\left( W_{n}\right) _{l-j}\left(
W_{n}^{m+1}\right) _{-kl}=\left( W_{n}^{m+1}\right) _{-k-j},
\end{gather*}
Thus, $G_{jk}(z)=G_{-k-j}(z)$.
\begin{gather*}
2. \quad \left( W_{n}^{m+1}\right) _{jk} =\sum_{l=-n}^{n}\left(
W_{n}\right) _{jl}\left( W_{n}^{m+1}\right)
_{lk}=\sum_{l=-n}^{n}\left( W_{n}\right)
_{j-l}\left( W_{n}^{m+1}\right) _{-lk} \\
\phantom{2. \quad  \left( W_{n}^{m+1}\right) _{jk}}{}
=\sum_{l=-n}^{n}\left( W_{n}\right) _{-jl}\left(
W_{n}^{m+1}\right) _{l-k}=\left( W_{n}^{m+1}\right) _{-j-k},
\end{gather*}
Thus, $G_{jk}(z)=G_{-j-k}(z)$.
\begin{gather*}
3. \quad \left( W_{n}^{m+1}\right) _{jk} =\sum_{l=-n}^{n}\left(
W_{n}\right)
_{jl}\left( W_{n}^{m+1}\right) _{lk} \\
\phantom{3. \quad \left( W_{n}^{m+1}\right) _{jk}}{}
=\sum_{l=-n}^{n}\left( W_{n}\right) _{-jl}\left(
W_{n}^{m+1}\right) _{lk}=\left( W_{n}^{m+1}\right) _{-jk}.
\end{gather*}
Thus,
\begin{gather}
G_{jk}(z)=G_{-jk}(z)-z^{-1}(\delta _{jk}-\delta _{-jk})
\label{GMe}
\end{gather}
and, hence,
\[
g(z)=\hat{g}(z)-z^{-1}, \qquad {\rm where} \qquad
\hat{g}(z)=\sum\limits_{r=-n}^{n}G_{r-r}(z).
\]

Unfortunately, there is no any such property for the fourth
ensemble.

Now using the resolvent identity for the average $\left\langle
G_{pq}(z)\right\rangle $, relation (\ref{NF}) and formula for the
derivative of the resolvent
\begin{gather}
G^{\prime }(z)\cdot X=-G(z)XG(z),  \label{RD}
\end{gather}
we obtain
\begin{gather*}
\left\langle G_{pq}(z)\right\rangle =-z^{-1}\delta
_{pq}+z^{-1}\left\langle \left( W_{n}G(z)\right) _{pq}\right\rangle  \\
\phantom{\left\langle G_{pq}(z)\right\rangle }{} =-z^{-1}\delta
_{pq}+z^{-1}\frac{1}{\sqrt{2n}}\sum\limits_{r=-n}^{n}\left \langle
\frac{1}{2}\left( \frac{\partial }{\partial \xi
_{pr}}+i\frac{\partial }
{\partial \eta _{pr}}\right) G_{rq}(z)\right\rangle  \\
\phantom{\left\langle G_{pq}(z)\right\rangle }{}
=-z^{-1}\delta _{pq}+z^{-1}\frac{1}{\sqrt{2n}}\sum\limits_{r,j,k=-n}^{n}%
\left\langle G_{rj}(z)\left( W_{n}^{\prime }\right)
_{jk}G_{kq}(z)\right\rangle ,
\end{gather*}
where $W_{n}^{\prime }=\frac{1}{2}\left( \frac{\partial }{\partial \xi _{pr}}+i\frac{%
\partial }{\partial \eta _{pr}}\right) W_{n}$. Now we calculate
$W_{n}^{\prime }$ for all four ensembles:
\begin{gather*}
1. \quad W_{n}^{\prime } =\frac{1}{2\sqrt{2n}}\left(
\begin{array}{c}
\delta _{jp}\delta _{kr}+\delta _{jr}\delta _{kp}+\delta
_{j-r}\delta
_{k-p}+\delta _{j-p}\delta _{k-r} \\
-\delta _{jp}\delta _{kr}+\delta _{jr}\delta _{kp}-\delta
_{j-r}\delta
_{k-p}+\delta _{j-p}\delta _{k-r}%
\end{array}
\right) \\
\phantom{1. \quad W_{n}^{\prime }}{} =\frac{1}{\sqrt{2n}}\left(
\delta _{jr}\delta _{kp}+\delta _{j-p}\delta _{k-r}\right) ;
\\
2. \quad W_{n}^{\prime } =\frac{1}{2\sqrt{2n}}\left(
\begin{array}{c}
\delta _{jp}\delta _{kr}+\delta _{jr}\delta _{kp}+\delta
_{j-p}\delta
_{k-r}+\delta _{j-r}\delta _{k-p} \\
-\delta _{jp}\delta _{kr}+\delta _{jr}\delta _{kp}-\delta
_{j-p}\delta _{k-r}+\delta _{j-r}\delta _{k-p}
\end{array}
\right) \\
\phantom{2. \quad W_{n}^{\prime }}{} =\frac{1}{\sqrt{2n}}\left(
\delta _{jr}\delta _{kp}+\delta _{j-r}\delta _{k-p}\right) ;
\\
3. \quad W_{n}^{\prime } =\frac{1}{2\sqrt{2n}}\left(
\begin{array}{c}
\delta _{jp}\delta _{kr}+\delta _{jr}\delta _{kp}+\delta
_{j-p}\delta
_{kr}+\delta _{jr}\delta _{k-p} \\
-\delta _{jp}\delta _{kr}+\delta _{jr}\delta _{kp}-\delta
_{j-p}\delta _{kr}+\delta _{jr}\delta _{k-p}
\end{array}
\right) \\
\phantom{3. \quad  W_{n}^{\prime }}{} =\frac{1}{\sqrt{2n}}\left(
\delta _{jr}\delta _{kp}+\delta _{jr}\delta _{k-p}\right) ;
\\
4. \quad W_{n}^{\prime } =\frac{1}{2\sqrt{2n}}\left(
\begin{array}{c}
\delta _{jp}\delta _{kr}+\delta _{jr}\delta _{kp}+\delta
_{jr}\delta
_{k-p}+\delta _{j-p}\delta _{kr} \\
-\delta _{jp}\delta _{kr}+\delta _{jr}\delta _{kp}-\delta
_{jr}\delta
_{k-p}+\delta _{j-p}\delta _{kr}%
\end{array}
\right) \\
\phantom{4. \quad W_{n}^{\prime }}{} =\frac{1}{\sqrt{2n}}\left(
\delta _{jr}\delta _{kp}+\delta _{j-p}\delta _{kr}\right) .
\end{gather*}

Using these formulas, we obtain the following relations:
\begin{gather}
1. \quad \left\langle G_{pq}(z)\right\rangle =-z^{-1}\delta
_{pq}-z^{-1}\left\langle
g(z)G_{pq}(z)\right\rangle -z^{-1}\left\langle \frac{1}{2n}%
\sum\limits_{r=-n}^{n}G_{r-p}(z)G_{-rq}(z)\right\rangle ;
\nonumber\\
2. \quad \left\langle G_{pq}(z)\right\rangle =-z^{-1}\delta
_{pq}-z^{-1}\left\langle
g(z)G_{pq}(z)\right\rangle -z^{-1}\left\langle \hat{g}(z)G_{-pq}(z)\right%
\rangle ;
\nonumber\\
3. \quad \left\langle G_{pq}(z)\right\rangle =-z^{-1}\delta
_{pq}-z^{-1}\left\langle g(z)G_{pq}(z)\right\rangle
-z^{-1}\left\langle g(z)G_{-pq}(z)\right\rangle ;
\nonumber\\
4. \quad \left\langle G_{pq}(z)\right\rangle =-z^{-1}\delta
_{pq}-z^{-1}\left\langle g(z)G_{pq}(z)\right\rangle
-z^{-1}\left\langle \frac{1}{2n}
\sum\limits_{r=-n}^{n}G_{r-p}(z)G_{rq}(z)\right\rangle .
\label{Gpg5}
\end{gather}

Now we put $p=q$ in all four cases and $p=-q$ another time in the
second case, and apply $\frac{1}{2n}\sum\limits_{p=-n}^{n}$. Thus,
using also the additional symmetries of the resolvents of
ensembles (\ref{Bel})--(\ref{Me}), we obtain:
\begin{gather*}
1. \quad \left\langle g(z)\right\rangle =-z^{-1}\left(
1+\left\langle
g(z)\right\rangle ^{2}\right) -z^{-1}\left[ \frac{1}{2n}\left\langle \frac{1%
}{2n}\mathrm{Tr}G^{2}(z)\right\rangle +\left\langle g^{\circ
}(z)g(z)\right\rangle \right] ,
\end{gather*}
where $g^{\circ }(z)=g(z)-\left\langle g(z)\right\rangle $;
\begin{gather}
2. \quad \left\langle g(z)\right\rangle  =  -z^{-1}\left(
1+\left\langle g(z)\right\rangle ^{2}+\left\langle
\hat{g}^{2}(z)\right\rangle \right)
 -z^{-1}\left[ \left\langle g^{\circ }(z)g(z)\right\rangle +\left\langle
\hat{g}^{\circ }(z)\hat{g}(z)\right\rangle \right] , \nonumber\\
\phantom{2. \quad{}}{} \left\langle \hat{g}(z)\right\rangle  =
-2z^{-1}\left\langle g(z)\right\rangle \left\langle
\hat{g}(z)\right\rangle -2z^{-1}\left\langle
g^{\circ }(z)\hat{g}(z)\right\rangle ;\nonumber\\
3. \quad \left\langle g(z)\right\rangle  =  -z^{-1}\left(
1+\left\langle
g(z)\right\rangle ^{2}+\left\langle g(z)\right\rangle \left\langle \hat{g}%
(z)\right\rangle \right) -z^{-1}\left[ \left\langle g^{\circ
}(z)g(z)\right\rangle +\left\langle
\hat{g}^{\circ }(z)\hat{g}(z)\right\rangle \right] , \nonumber\\
\phantom{3. \quad{}}{} \hat{g}(z)  =  g(z)+z^{-1};  \label{g3}
\\
4. \quad \left\langle g(z)\right\rangle =-z^{-1}\left(
1+\left\langle
g(z)\right\rangle ^{2}\right) -z^{-1}\left[ \frac{1}{2n}\left\langle \frac{1%
}{2n}\,\mathrm{Tr}\, P(z)G(z)\right\rangle +\left\langle g^{\circ
}(z)g(z)\right\rangle \right] ,\nonumber
\end{gather}
where matrix $P(z)$ is defined by $P_{xy}(z)=G_{y-x}(z).$

In the appendix we prove that the variances of random variables
$g(z)$ in all cases above are of the order $O(n^{-2})$ uniformly
in $z$ for some compact in $C_{\pm }$ (as well as the variance of
$\hat{g}(z)$ in the second case). Besides, using Schwartz
inequality for the matrix scalar product $ (A,B)=\mathrm{Tr}\,AB$,
we obtain
\begin{gather*}
\left\vert \frac{1}{2n}\,\mathrm{Tr}\,P(z)G(z)\right\vert \leq
\left( \frac{1}{ 2n}\,\mathrm{Tr}\,P(z)P^{\dagger }(z)\right)
^{1/2}\left( \frac{1}{2n}\,\mathrm{Tr} G(z)G^{\dagger }(z)\right)
^{1/2}\leq \frac{1}{\left\vert \mathrm{Im\,}
z\right\vert ^{2}}, \\
\left\vert \frac{1}{2n}\,\mathrm{Tr}\,G^{2}(z)\right\vert
\leq\frac{1}{ \left\vert \mathrm{Im\,}z\right\vert ^{2}}.
\end{gather*}
Thus, all terms in square brackets in all four cases are at least
of the order $O(n^{-1})$. Hence, in the first and in the fourth
cases we obtain the following limiting equation:
\begin{gather}
f(z)=-z^{-1}\left( 1+f^{2}(z)\right) ,  \label{fsc}
\end{gather}
which is the equation for $f_{sc}(z)$ --- the Stieltjes transform
of the semi-circle Law.

Besides, since
\begin{gather*}
g(z) =-z_{pq}^{-1}-z^{-1}\frac{1}{2n}\,\mathrm{Tr}\,\left(
W_{n}G(z)\right) ,
\\
\left| \frac{1}{2n}\,\mathrm{Tr}\,\left( W_{n}G(z)\right) \right|
\leq \frac{1 }{\left| \mathrm{Im\,}z\right| }\left(
\frac{1}{2n}\,\mathrm{Tr}\,
W_{n}W_{n}^{\dagger }\right) ^{1/2}, \\
\left\langle \left( \frac{1}{2n}\,\mathrm{Tr}\,W_{n}W_{n}^{\dagger
}\right) ^{1/2}\right\rangle \leq \left\langle
\frac{1}{2n}\,\mathrm{Tr}\, W_{n}W_{n}^{\dagger }\right\rangle
^{1/2}\leq 1,
\end{gather*}
then for all $z$ with e.g.\ $\left| \mathrm{Im\,}z\right| \geq 3$
uniformly in $n$ we have in all cases
\begin{gather}
\left| 1+2z^{-1}\left\langle g(z)\right\rangle \right|
>\frac{1}{2}. \label{1p2g}
\end{gather}
{\samepage Thus, in the second case $\left\langle \hat{g}(z)\right\rangle $
of the order $O(n^{-2})$:
\[
\left\langle \hat{g}(z)\right\rangle =-2z^{-1}\left(
1+2z^{-1}\left\langle g(z)\right\rangle \right) ^{-1}\left\langle
g^{\circ }(z)\hat{g} (z)\right\rangle .
\]
Hence, the second case lead to the same limiting equation
(\ref{fsc}).}

As for the third case, it leads to the following equation
\begin{gather}
f(z)=-z^{-1}\left( 1+2f^{2}(z)+z^{-1}f(z)\right) .  \label{Meeq}
\end{gather}
Its solution in the class of Nevanlinna functions is the Stieltjes
transform of the measure (\ref{MeN}).

The convergence with probability one in all four cases follows
from the bounds for the variances in the section bellow and the
Borel--Cantelli lemma.

\section{The correlation functions}

As in the previous section, we perform our calculations in
parallel for all four ensembles.

Using the resolvent identity for the average $\left\langle
g^{\circ }(z_{1})G_{pq}(z_{2})\right\rangle $, relations
(\ref{NF}) and (\ref{RD}), we obtain
\begin{gather*}
\left\langle g^{\circ }(z_{1})G_{pq}(z_{2})\right\rangle
=z_{2}^{-1}\frac{1
}{\sqrt{2n}}\sum\limits_{r,j,k=-n}^{n}\left\langle g^{\circ
}(z_{1})G_{rj}(z_{2})\left( W_{n}^{\prime }\right)
_{jk}G_{kq}(z_{2})\right\rangle \\
\phantom{\left\langle g^{\circ
}(z_{1})G_{pq}(z_{2})\right\rangle=}{}
+z_{2}^{-1}\frac{1}{\left( 2n\right) ^{3/2}}\sum\limits_{l.r,j,k=-n}^{n}%
\left\langle G_{lj}(z_{1})\left( W_{n}^{\prime }\right)
_{jk}G_{kl}(z_{1})G_{rq}(z_{2})\right\rangle .
\end{gather*}
Substituting in this relation the value of $W_{n}^{\prime }$ in
all four cases and using the symmetries of the resolvents, we
obtain
\begin{gather*}
1. \quad \left\langle g^{\circ }(z_{1})G_{pq}(z_{2})\right\rangle
=-z_{2}^{-1}\left\langle g^{\circ
}(z_{1})g(z_{2})G_{pq}(z_{2})\right\rangle -z_{2}^{-1}\left\langle
g^{\circ
}(z_{1})\frac{1}{2n}G_{pq}^{2}(z_{2})\right\rangle \\
\phantom{1. \quad \left\langle g^{\circ
}(z_{1})G_{pq}(z_{2})\right\rangle=}{} -z_{2}^{-1}\frac{1}{\left(
2n\right) ^{2}}\left( \left\langle \left(
G^{2}(z_{1})G(z_{2})\right) _{pq}\right\rangle +\left\langle
\left( G(z_{2})G^{2}(z_{1})\right) _{-q-p}\right\rangle \right) ;
\\
2. \quad \left\langle g^{\circ }(z_{1})G_{pq}(z_{2})\right\rangle
= -z_{2}^{-1}\left\langle g^{\circ
}(z_{1})g(z_{2})G_{pq}(z_{2})\right\rangle -z_{2}^{-1}\left\langle
g^{\circ
}(z_{1})\hat{g}(z_{2})G_{-pq}(z_{2})\right\rangle \\
\phantom{2. \quad \left\langle g^{\circ
}(z_{1})G_{pq}(z_{2})\right\rangle =}{} -z_{2}^{-1}\frac{1}{\left(
2n\right) ^{2}}\left( \left\langle \left(
G^{2}(z_{1})G(z_{2})\right) _{pq}\right\rangle +\left\langle
\left( G(z_{2})G^{2}(z_{1})\right) _{-p-q}\right\rangle \right) ;
\\
3. \quad \left\langle g^{\circ }(z_{1})G_{pq}(z_{2})\right\rangle
= -z_{2}^{-1}\left\langle g^{\circ
}(z_{1})g(z_{2})G_{pq}(z_{2})\right\rangle -z_{2}^{-1}\left\langle
g^{\circ
}(z_{1})g(z_{2})G_{-pq}(z_{2})\right\rangle \\
\phantom{3. \quad \left\langle g^{\circ
}(z_{1})G_{pq}(z_{2})\right\rangle =}{} -z_{2}^{-1}\frac{1}{\left(
2n\right) ^{2}}\left( \left\langle \left(
G^{2}(z_{1})G(z_{2})\right) _{pq}\right\rangle +\left\langle
\left( G^{2}(z_{1})G(z_{2})\right) _{-pq}\right\rangle \right) ;
\\
4. \quad \left\langle g^{\circ }(z_{1})G_{pq}(z_{2})\right\rangle
= -z_{2}^{-1}\left\langle g^{\circ
}(z_{1})g(z_{2})G_{pq}(z_{2})\right\rangle -z_{2}^{-1}\left\langle
g^{\circ
}(z_{1})\frac{1}{2n}\sum\limits_{r=-n}^{n}G_{r-p}(z_{2})G_{rq}(z_{2})
\right\rangle \!\!\!\\
\phantom{4. \quad \left\langle g^{\circ
}(z_{1})G_{pq}(z_{2})\right\rangle =}{} -z_{2}^{-1}\frac{1}{\left(
2n\right) ^{2}}\left( \left\langle \left(
G^{2}(z_{1})G(z_{2})\right) _{pq}\right\rangle +\left\langle
\sum\limits_{r=-n}^{n}\left( G^{2}(z_{1})\right)
_{r-p}G_{rq}(z_{2})\right\rangle \right) .\!
\end{gather*}

Then we put $p=q$ in all four cases and $p=-q$ another time in the
second case, and apply $\frac{1}{2n}\sum\limits_{p=-n}^{n}$ and
obtain
\begin{gather}
1. \quad \left\langle g^{\circ }(z_{1})g(z_{2})\right\rangle
=-2z_{2}^{-1}\left\langle g(z_{2})\right\rangle \left\langle
g^{\circ
}(z_{1})g(z_{2})\right\rangle  \nonumber \\
\phantom{1. \quad \left\langle g^{\circ
}(z_{1})g(z_{2})\right\rangle=}{} -z_{2}^{-1}\frac{2}{\left(
2n\right) ^{2}}\left\langle \frac{1}{2n}\, \mathrm{Tr}\,
G^{2}(z_{1})G(z_{2})\right\rangle +r_{1,n},  \label{gv1}
\end{gather}
where
\begin{gather}
r_{1,n}=-z_{2}^{-1}\left( \left\langle g^{\circ }(z_{1})\left(
g^{\circ
}(z_{2})\right) ^{2}\right\rangle +\frac{1}{2n}\left\langle g^{\circ }(z_{1})%
\frac{1}{2n}\, \mathrm{Tr}\, G^{2}(z_{2})\right\rangle \right) ;
\label{r1}
\\
2. \quad \left\langle g^{\circ }(z_{1})g(z_{2})\right\rangle  =
 -2z_{2}^{-1}\left(
\left\langle g(z_{2})\right\rangle \left\langle g^{\circ
}(z_{1})g(z_{2})\right\rangle +\left\langle
\hat{g}(z_{2})\right\rangle
\left\langle g^{\circ }(z_{1})\hat{g}(z_{2})\right\rangle \right) \nonumber\\
\phantom{2. \quad \left\langle g^{\circ
}(z_{1})g(z_{2})\right\rangle  = }{}
 -z_{2}^{-1}\frac{2}{\left( 2n\right) ^{2}}\left\langle \frac{1}{2n}\,
\mathrm{Tr}\,G^{2}(z_{1})G(z_{2})\right\rangle +r_{2,n}, \nonumber\\
\phantom{2. \quad{}}{} \left\langle \hat{g}^{\circ
}(z_{1})\hat{g}(z_{2})\right\rangle  = -2z_{2}^{-1}\left(
\left\langle g(z_{2})\right\rangle \left\langle \hat{g} ^{\circ
}(z_{1})\hat{g}(z_{2})\right\rangle +\left\langle \hat{g}
(z_{2})\right\rangle \left\langle g^{\circ }(z_{1})\hat{g}
(z_{2})\right\rangle \right) \nonumber\\
\phantom{2. \quad \left\langle g^{\circ
}(z_{1})g(z_{2})\right\rangle  =}{}
 -z_{2}^{-1}\frac{2}{\left( 2n\right) ^{2}}\left\langle \frac{1}{2n}\,
\mathrm{Tr}\,G^{2}(z_{1})G(z_{2})\right\rangle +r_{3,n},\nonumber
\end{gather}
where
\begin{gather*}
r_{2,n} =-z_{2}^{-1}\left( \left\langle g^{\circ }(z_{1})\left(
g^{\circ
}(z_{2})\right) ^{2}\right\rangle +\left\langle g^{\circ }(z_{1})\left( \hat{%
g}^{\circ }(z_{2})\right) ^{2}\right\rangle \right) , \\
r_{3,n} =-2z_{2}^{-1}\left\langle \hat{g}^{\circ
}(z_{1})\hat{g}^{\circ }(z_{2})g^{\circ }(z_{2})\right\rangle ;
\\
3. \quad \left\langle g^{\circ }(z_{1})g(z_{2})\right\rangle =
-4z_{2}^{-1}\left\langle g(z_{2})\right\rangle \left\langle
g^{\circ }(z_{1})g(z_{2})\right\rangle -z_{2}^{-2}\left\langle
g^{\circ
}(z_{1})g(z_{2})\right\rangle \\
\phantom{3. \quad \left\langle g^{\circ
}(z_{1})g(z_{2})\right\rangle =}{} -z_{2}^{-1}\frac{1}{\left(
2n\right) ^{2}}\left\langle \frac{1}{2n}
\, \mathrm{Tr}\, G^{2}(z_{1})G(z_{2})\right\rangle \\
\phantom{3. \quad \left\langle g^{\circ
}(z_{1})g(z_{2})\right\rangle =}{} -z_{2}^{-1}\frac{1}{\left(
2n\right) ^{2}}\left\langle \frac{1}{2n}
\sum\limits_{p=-n}^{n}\left( G^{2}(z_{1})G(z_{2})\right)
_{-pp}\right\rangle +r_{4,n},
\end{gather*}
where
\begin{gather*}
r_{4,n}=-2z_{2}^{-1}\left\langle g^{\circ }(z_{1})\left( g^{\circ
}(z_{2})\right) ^{2}\right\rangle ;
\\
4. \quad \left\langle g^{\circ }(z_{1})g(z_{2})\right\rangle =
-2z_{2}^{-1}\left\langle g(z_{2})\right\rangle \left\langle
g^{\circ
}(z_{1})g(z_{2})\right\rangle \\
\phantom{4. \quad \left\langle g^{\circ
}(z_{1})g(z_{2})\right\rangle =}{} -z_{2}^{-1}\frac{1}{\left(
2n\right) ^{2}}\left\langle \frac{1}{2n}\, \mathrm{Tr}\,
G^{2}(z_{1})G(z_{2})\right\rangle +r_{5,n},
\end{gather*}
where
\begin{gather}
r_{5,n} =-z_{2}^{-1}\frac{1}{2n}\left\langle g^{\circ
}(z_{1})\frac{1}{2n}\, \mathrm{Tr}\,P(z_{2})G(z_{2})\right\rangle
-z_{2}^{-1}\left\langle g^{\circ }(z_{1})\left( g^{\circ
}(z_{2})\right)
^{2}\right\rangle  \nonumber \\
\phantom{r_{5,n} =}{} -z_{2}^{-1}\frac{1}{\left( 2n\right)
^{2}}\left\langle \frac{1}{2n} \sum\limits_{r,p=-n}^{n}\left(
G^{2}(z_{1})\right) _{r-p}G_{rp}(z_{2})\right\rangle .  \label{r5}
\end{gather}

As we show in the appendix, all $r_{j,n}$, $j=1,\ldots ,5$ are of
the order
$o(n^{-2})$. Thus, as one can easily show, all correlation functions $%
F(z_{1},z_{2})=\left\langle g^{\circ }(z_{1})g(z_{2})\right\rangle
$ above
are of the order $O(n^{-2})$. Moreover, since $\left\langle \hat{g}%
(z_{2})\right\rangle $ is of the order $O(n^{-2})$ in the second
case, its
easy to see that cases one and two lead to the same relation for $%
F(z_{1},z_{2})$
\begin{gather}
F(z_{1},z_{2})=-2z_{2}^{-1}\left\langle g(z_{2})\right\rangle
F(z_{1},z_{2})-z_{2}^{-1}\frac{2}{\left( 2n\right)
^{2}}\left\langle
\frac{1}{2n}\,\mathrm{Tr}\,G^{2}(z_{1})G(z_{2})\right\rangle
+o\left(n^{-2}\right). \label{F12}
\end{gather} As to the case four, it leads to
\begin{gather}
F(z_{1},z_{2})=-2z_{2}^{-1}\left\langle g(z_{2})\right\rangle
F(z_{1},z_{2})-z_{2}^{-1}\frac{1}{\left( 2n\right)
^{2}}\left\langle \frac{1}{2n}\,
\mathrm{Tr}\,G^{2}(z_{1})G(z_{2})\right\rangle
+o\left(n^{-2}\right). \label{F4}
\end{gather}
Besides, due to the resolvent identity we have
\begin{gather}
\frac{1}{2n}\,\mathrm{Tr}\,G^{2}(z_{1})G(z_{2})=\frac{1}{z_{1}-z_{2}}\left(
\frac{1}{2n}\,\mathrm{Tr}\,
G^{2}(z_{1})-\frac{g(z_{1})-g(z_{2})}{z_{1}-z_{2}}\right) .
\label{G1G2}
\end{gather}
In addition, as we show in the appendix, in these cases
\begin{gather}
\left\langle \frac{1}{2n}\,\mathrm{Tr}\,G^{2}(z)\right\rangle =
\frac{\left\langle g(z)\right\rangle }{1-\left\langle
g(z)\right\rangle ^{2}} +O\left(n^{-1}\right). \label{G2}
\end{gather}
Thus, substituting in the relations (\ref{F12}), (\ref{F4}) the
expressions (\ref{G1G2}), (\ref{G2}) and using the
equation~(\ref{fsc}) for the limit of $\langle g(z)\rangle$, we
obtain in the cases one and two the GOE correlator
asymp\-totic~(\ref{CGOE}) and in the case four the twice less GUE
asymptotic.

To treat the third case we use (\ref{GMe}) and obtain that
\[
\frac{1}{2n}\sum\limits_{p=-n}^{n}\left(
G^{2}(z_{1})G(z_{2})\right) _{-pp}=
\frac{1}{2n}\mathrm{Tr}G^{2}(z_{1})G(z_{2})+\frac{1}{z_{1}^{2}z_{2}}.
\]
This gives the following relation for $F(z_{1},z_{2})$
\begin{gather}
F(z_{1},z_{2}) =\left( -4\frac{\left\langle g(z_{2})\right\rangle }{z_{2}}%
-z_{2}^{-2}\right) F(z_{1},z_{2})  \nonumber\\
\phantom{F(z_{1},z_{2}) =}{}-\frac{z_{2}^{-1}}{\left( 2n\right) ^{2}}\left( \frac{1}{z_{1}^{2}z_{2}}+%
\frac{1}{z_{1}-z_{2}}\left\langle \frac{1}{2n}\,\mathrm{Tr}\,G^{2}(z_{1})-\frac{%
g(z_{1})-g(z_{2})}{z_{1}-z_{2}}\right\rangle \right)
+o\left(n^{-2}\right).  \label{FMe}
\end{gather}
We show also in the appendix that in this case
\[
\left\langle \frac{1}{2n}\,\mathrm{Tr}\, G^{2}(z)\right\rangle =-
\frac{\left\langle g(z)\right\rangle
}{z}\frac{1-z^{-2}}{1+\left\langle g(z)\right\rangle
z^{-1}+z^{-2}}+o\left(n^{-2}\right).
\]
Substituting this relation in (\ref{FMe}) we obtain
\begin{gather*}
F(z_{1},z_{2})=\frac{1}{n^2}\left(-\frac{\frac{1}{\left(
z_{1}z_{2}\right) ^{2}}+\frac{2}{z_{1}}
\frac{f(z_{1})-f(z_{2})}{\left( z_{1}-z_{2}\right) ^{2}}}
{1+z_{1}^{-2}+4z_{1}^{-1}f(z_{1})}\right.\\
\left.\phantom{F(z_{1},z_{2})=}{}
-2\frac{1-z_{2}^{-2}}{z_{1}-z_{2}}
\frac{f(z_{2})}{z_{1}z_{2}}\prod\limits_{k=1,2}\left(
1+z_{k}^{-2}+4\frac{f(z_{k})
}{z_{k}}\right)\right)+o\left(n^{-2}\right).
\end{gather*}
Then, using the equation (\ref{Meeq}), we rewrite this relation in
the form (\ref{MeC}).

\section{Conclusion}

The purpose of this paper was to answer the question: ``Can the 
additional symmetry properties influence on the asymptotic behavior 
of eigenvalue distribution of GUE?'' The negative answer for the 
three cases of additional symmetry is not surprising, as these 
symmetries leave the number of independent random parameters of the 
order $n^2$. The effect when in one case the additional symmetry 
essentially changes the limiting eigenvalue counting measure is very 
unexpected, especially the appearance of the gap in the support of 
limiting NCM. Unfortunately, the physical application of this effect 
is unknown to the author, though one of the other considered 
ensembles (flip matrix model) was used as an approach to weak 
coupling regime of the Anderson model.

\appendix
\section{Appendix}
\begin{proposition}
The variance $v=\left\langle \left| g^{\circ }(z)\right|
^{2}\right\rangle $
is of the order $O(n^{-2})$ in all four cases, and the terms $r_{j,n}$, $%
j=1,\ldots ,5$ are of the order $o(n^{-2})$.
\end{proposition}

\begin{proof} First we proof that the variance is of the order
$O(n^{-2})$
in all four cases. Indeed, in the first case, using (\ref{gv1}) with $z_{2}=%
\overline{z_{1}}=z,$ we obtain
\[
v(1+2z^{-1}\left\langle g(z)\right\rangle
)=-z_{2}^{-1}\frac{2}{\left(
2n\right) ^{2}}\left\langle \frac{1}{2n}\,\mathrm{Tr}\, G^{2}(z_{1})G(z_{2})%
\right\rangle +r_{1,n}.
\]%
Besides, using the Schwartz inequality we obtain from (\ref{r1})
\[
r_{1,n}\leq |z|^{-1}\left( \frac{1}{\left\vert
\mathrm{Im\,}z\right\vert }v+ \frac{1}{2n\left\vert
\mathrm{Im\,}z\right\vert ^{2}}v^{1/2}\right) .
\]
Thus, due to the bounds (\ref{1p2g}) and
\begin{gather}
\left\vert
\frac{1}{2n}\,\mathrm{Tr}\,G^{2}(z_{1})G(z_{2})\right\vert \leq
\frac{1}{\left\vert \mathrm{Im\,}z\right\vert ^{3}},  \label{G2G}
\end{gather}
we have for $\left\vert \mathrm{Im\,}z\right\vert \geq 3$ the
inequality
\[
v\leq \frac{2}{9\left( 2n\right) ^{2}}+\frac{1}{2n}v^{1/2},
\]
which leads to $v=O(n^{-2})$. For the other cases the proofs are
analogous.

To prove $r_{1,n}=o(n^{-2})$ for
\[
r_{1,n}=-z_{2}^{-1}\left( \left\langle g^{\circ }(z_{1})\left(
g^{\circ }(z_{2})\right) ^{2}\right\rangle
+\frac{1}{2n}\left\langle g^{\circ }(z_{1})
\frac{1}{2n}\,\mathrm{Tr}\, G^{2}(z_{2})\right\rangle \right) ,
\]
we rewrite the second term in the parentheses as
\[
\frac{1}{2n}\left\langle g^{\circ
}(z_{1})\frac{1}{2n}\,\mathrm{Tr}\, G^{2}(z_{2})\right\rangle
=\frac{1}{2n}\left\langle g^{\circ }(z_{1})\frac{1}{2n} \,
\mathrm{Tr}\, \frac{\partial }{\partial
z_{2}}G(z_{2})\right\rangle =\frac{1}{2n}\frac{\partial }{\partial
z_{2}}\left\langle g^{\circ }(z_{1})g^{\circ }(z_{2})\right\rangle
.
\]
Since the value $\left\langle g^{\circ }(z_{1})g^{\circ
}(z_{2})\right\rangle $ is analytical and uniformly in $n$ bounded
for $\left\vert \mathrm{Im\,}z_{1,2}\right\vert \geq 3$, and
since, due to the Schwartz inequality $\left\vert \left\langle
g^{\circ }(z_{1})g^{\circ }(z_{2})\right\rangle \right\vert \leq
v=O(n^{-2})$, its derivative on $z_{2} $ is also of the order
$O(n^{-2})$ and hence the second term is of the order $O(n^{-3})$.

To prove that the first term is $o(n^{-2})$ let us consider
\[
\left\langle \left\vert g^{\circ }(z)\right\vert ^{4}\right\rangle
=\left\langle \left( g^{\circ }(z_{1})g^{\circ }(z_{2})\right)
^{2}\right\rangle =\left\langle R^{\circ }g(z_{2})\right\rangle
,\qquad R\equiv \left( g^{\circ }(z_{1})\right) ^{2}g^{\circ
}(z_{2}),\qquad z_{1}=\overline{z_{2}}=z.
\]
Then, using the resolvent identity for the average $\left\langle
R^{\circ }G_{pq}(z_{2})\right\rangle $, relations (\ref{NF}) and
(\ref{RD}), we obtain
\begin{gather*}
\left\langle R^{\circ }G_{pq}(z_{2})\right\rangle=
z_{2}^{-1}\frac{1}{\sqrt{2n}}\sum\limits_{r,j,k=-n}^{n}\left\langle
R^{\circ }G_{rj}(z_{2})\left( W_{n}^{\prime }\right)
_{jk}G_{kq}(z_{2})\right\rangle
\\
\phantom{\left\langle R^{\circ }G_{pq}(z_{2})\right\rangle=}{}
+z_{2}^{-1}\frac{2}{\left( 2n\right)
^{3/2}}\sum\limits_{l.r,j,k=-n}^{n} \left\langle g^{\circ
}(z_{1})g^{\circ }(z_{2})G_{lj}(z_{1})\left(
W_{n}^{\prime }\right) _{jk}G_{kl}(z_{1})G_{rq}(z_{2})\right\rangle \\
\phantom{\left\langle R^{\circ }G_{pq}(z_{2})\right\rangle=}{}
+z_{2}^{-1}\frac{1}{\left( 2n\right)
^{3/2}}\sum\limits_{l.r,j,k=-n}^{n} \left\langle \left( g^{\circ
}(z_{1})\right) ^{2}G_{lj}(z_{2})\left( W_{n}^{\prime }\right)
_{jk}G_{kl}(z_{2})G_{rq}(z_{2})\right\rangle .
\end{gather*}
Substituting in this relation the value of $W_{n}^{\prime }$ and
using the symmetry of the resolvent we obtain
\begin{gather*}
\left\langle R^{\circ }G_{pq}(z_{2})\right\rangle
=-z_{2}^{-1}\left\langle R^{\circ
}g(z_{2})G_{pq}(z_{2})\right\rangle -z_{2}^{-1}\left\langle
R^{\circ }\frac{1}{2n}G_{pq}^{2}(z_{2})\right\rangle \\
\phantom{\left\langle R^{\circ }G_{pq}(z_{2})\right\rangle =}{}
-z_{2}^{-1}\frac{2}{\left( 2n\right) ^{2}}\left\langle g^{\circ
}(z_{1})g^{\circ }(z_{2})\left( \left( G^{2}(z_{1})G(z_{2})\right)
_{pq}+\left( G(z_{2})G^{2}(z_{1})\right) _{-q-p}\right) \right\rangle \\
\phantom{\left\langle R^{\circ }G_{pq}(z_{2})\right\rangle =}{}
-z_{2}^{-1}\frac{2}{\left( 2n\right) ^{2}}\left\langle \left(
g^{\circ }(z_{1})\right) ^{2}G_{pq}^{3}(z_{2})\right\rangle .
\end{gather*}
Then we put $p=q$ in all four cases and $p=-q$ another time in the
second case, and apply $\frac{1}{2n}\sum\limits_{p=-n}^{n}$ and
obtain
\begin{gather*}
\left\langle \left\vert g^{\circ }(z)\right\vert
^{4}\right\rangle= \left\langle R^{\circ }g(z_{2})\right\rangle
=-z_{2}^{-1}\left\langle R^{\circ }g^{2}(z_{2})\right\rangle
-z_{2}^{-1}\frac{1}{2n}\left\langle
R^{\circ }\frac{1}{2n}\,\mathrm{Tr}\,G^{2}(z_{2})\right\rangle \\
\phantom{\left\langle \left\vert g^{\circ }(z)\right\vert
^{4}\right\rangle=}{} -z_{2}^{-1}\frac{4}{\left( 2n\right)
^{2}}\left\langle g^{\circ }(z_{1})g^{\circ
}(z_{2})\frac{1}{2n}\,\mathrm{Tr}\,\left(
G^{2}(z_{1})G(z_{2})\right) \right\rangle \\
\phantom{\left\langle \left\vert g^{\circ }(z)\right\vert
^{4}\right\rangle=}{} -z_{2}^{-1}\frac{2}{\left( 2n\right)
^{2}}\left\langle \left( g^{\circ }(z_{1})\right)
^{2}\frac{1}{2n}\,\mathrm{Tr}\,G^{3}(z_{2})\right\rangle .
\end{gather*}
Using this relation, the bounds (\ref{G2G}) and
\begin{gather*}
\left\vert \left\langle R^{\circ }g^{2}(z_{2})\right\rangle
\right\vert =\left\vert \left\langle R^{\circ }g^{\circ
}(z_{2})g(z_{2})\right\rangle +\left\langle R^{\circ
}g(z_{2})\right\rangle \left\langle g(z_{2})\right\rangle
\right\vert \leq 2\frac{\left\langle \left\vert g^{\circ
}(z)\right\vert ^{4}\right\rangle }{\left\vert \mathrm{Im\,}
z\right\vert },
\\
\left\vert \left\langle R^{\circ }\frac{1}{2n}\,
\mathrm{Tr}\,G^{2}(z_{2})\right\rangle \right\vert \leq
\frac{v}{\left\vert \mathrm{Im\,}z\right\vert ^{3}},
\end{gather*}
we obtain that for $\left\vert \mathrm{Im\,}z\right\vert \geq 3$ $%
\left\langle \left\vert g^{\circ }(z)\right\vert ^{4}\right\rangle
=O(n^{-3}) $. Thus, due to the Schwartz inequality the term
$\left\langle g^{\circ }(z_{1})\left( g^{\circ }(z_{2})\right)
^{2}\right\rangle $ is of the order $O(n^{-5/2})$ and, hence,
$r_{1,n}$ is of the same order.

The cases of the terms $r_{j,n}$, $j=2,\ldots ,5$ can be treated
analogously, with exception for the last term of $r_{5,n}$. The
last term of (\ref{r5})
\[
\frac{1}{\left( 2n\right) ^{2}}\left\langle \frac{1}{2n}\sum
\limits_{r,p=-n}^{n}\left( G^{2}(z_{1})\right)
_{r-p}G_{rp}(z_{2})\right\rangle
\]
can be treated as follows.

First, observe that in the case four $\left\langle
\hat{g}(z)\right\rangle =o(n^{-1})$. Indeed, using (\ref{Gpg5})
with $q=-p$, we obtain
\[
\left\langle \hat{g}(z)\right\rangle =-z^{-1}\left\langle
g(z)\right\rangle \left\langle \hat{g}(z)\right\rangle
-z^{-1}\left\langle g^{\circ }(z)\hat{g} (z)\right\rangle
-z^{-1}\frac{1}{2n}\left\langle \frac{1}{2n}\,\mathrm{Tr}\, \left(
G(z)G^{T}(z)\right) \right\rangle ,
\]
where $G^{T}$ is transpose of $G$. Due to the the Schwartz
inequality for
the trace, the last term in r.h.s.\ of this relation is of the order $%
O(n^{-1}) $. Since the variance of $g(z)$ is of the order
$O(n^{-2})$, the second term is at least of the order $O(n^{-1})$
(in fact it is of the order $O(n^{-2})$, since, as one can show,
the variance of $\hat{g}(z)$ is of the
same order). Thus, $\left\langle \hat{g}(z)\right\rangle $ is of the order $%
O(n^{-1})$. Its easy to show in the same way that
\[
\left\langle \hat{h}(z)\right\rangle =\left\langle \frac{1}{2n}%
\sum\limits_{j=-n}^{n}\left( G^{2}(z)\right) _{j-j}\right\rangle
\]
is also of the order $O(n^{-1})$ and its variance is of the order
$O(n^{-2})$.

Now, using the resolvent identity for the average of
\[
\Phi =\frac{1}{2n}\sum\limits_{p,q=-n}^{n}\left(
G^{2}(z_{1})\right) _{p-q}G_{pq}(z_{2}),
\]
relations (\ref{NF}) and (\ref{RD}), we obtain
\begin{gather*}
\left\langle \Phi \right\rangle = -z_{2}^{-1}\left\langle \hat{h}
(z_{1})\right\rangle -z_{2}^{-1}\frac{1}{\left( 2n\right) ^{3/2}}
\left\langle \sum\limits_{p,q,r,j,k=-n}^{n}\left(
G^{2}(z_{1})\right) _{p-q}G_{rj}(z_{2})\left( W_{n}^{\prime
}\right)
_{jk}G_{kq}(z_{2})\right\rangle \\
\phantom{\left\langle \Phi \right\rangle =}{}
-z_{2}^{-1}\frac{1}{\left( 2n\right) ^{3/2}}\left\langle
\sum\limits_{p,q,r,j,k,m=-n}^{n}G_{pj}(z_{1})\left( W_{n}^{\prime
}\right)
_{jk}G(z_{1})_{km}G_{m-q}(z_{1})G_{rq}(z_{2})\right\rangle \\
\phantom{\left\langle \Phi \right\rangle =}{}
-z_{2}^{-1}\frac{1}{\left( 2n\right) ^{3/2}}\left\langle
\sum\limits_{p,q,r,j,k,m=-n}^{n}G_{pm}(z_{1})G(z_{1})_{mj}\left(
W_{n}^{\prime }\right)
_{jk}G_{k-q}(z_{1})G_{rq}(z_{2})\right\rangle .
\end{gather*}
Substituting in this relation the value of $W_{n}^{\prime }$, we
obtain
\begin{gather*}
\left\langle \Phi \right\rangle = -z_{2}^{-1}\left\langle \hat{h}
(z_{1})\right\rangle -z_{2}^{-1}\left\langle g(z_{2})\Phi
\right\rangle -z_{2}^{-1}\frac{1}{2n}\left\langle
\frac{1}{2n}\sum\limits_{p,q,r=-n}^{n}
\left( G^{2}(z_{1})\right) _{p-q}G_{r-p}(z_{2})G_{rq}(z_{2})\right\rangle \\
\phantom{\left\langle \Phi \right\rangle =}{}
-z_{2}^{-1}\frac{1}{2n}\left\langle
\frac{1}{2n}\sum\limits_{p,q=-n}^{n}
\left( G(z_{1})G(z_{2})\right) _{pq}G^{2}{}_{p-q}(z_{1})\right\rangle \\
\phantom{\left\langle \Phi \right\rangle =}{}
-z_{2}^{-1}\frac{1}{2n}\left\langle
\frac{1}{2n}\sum\limits_{p,q=-n}^{n} \left(
G^{2}(z_{1})G(z_{2})\right) _{pq}G_{p-q}(z_{1})\right\rangle
-z_{2}^{-1}\left\langle \hat{g}(z_{1})\Phi \right\rangle
-z_{2}^{-1}\left\langle \hat{h}(z_{1})\Phi \right\rangle .
\end{gather*}
The first term of the r.h.s.\ is of the order $O(n^{-1})$, the
last five terms are of the same order, because of the Schwartz
inequality and of the bounds for the variances of $\hat{g}(z_{1})$
and $\hat{h}(z_{1})$. The second term we rewrite as follows
\[
-z_{2}^{-1}\left\langle g(z_{2})\Phi \right\rangle
=-z_{2}^{-1}\left\langle g(z_{2})\right\rangle \left\langle \Phi
\right\rangle -z_{2}^{-1}\left\langle g^{\circ }(z_{2})\Phi
\right\rangle ,
\]
where due to the Schwartz inequality the last term is also of the
order $O(n^{-1})$. Thus, we conclude that $\left\langle \Phi
\right\rangle $ is of the order $O(n^{-1})$ and, hence, the last
term of $r_{5,n}$ is of the order $ O(n^{-3})$.
\end{proof}

\begin{proposition}
In the third case we have
\[
\left\langle \frac{1}{2n}\,\mathrm{Tr}\,G^{2}(z)\right\rangle =
-\frac{\left\langle g(z)\right\rangle
}{z}\frac{1-z^{-2}}{1+\left\langle g(z)\right\rangle
z^{-1}+z^{-2}}+o\left(n^{-2}\right).
\]
\end{proposition}

\begin{proof} Indeed, we have
\[
\left\langle \frac{1}{2n}\,\mathrm{Tr}\,G^{2}(z)\right\rangle
=\left\langle
\frac{1}{2n}\,\mathrm{Tr}\,\frac{d}{dz}G(z)\right\rangle
=\frac{d}{dz} \left\langle g(z)\right\rangle .
\]
Hence, we can just take the derivative of the identity (\ref{g3})
for $\left\langle g(z)\right\rangle $. Since all terms of the
order $O(n^{-2})$ in square brackets in (\ref{g3}) are analytical
and uniformly bounded for $\left\vert
\mathrm{Im\,}z_{1,2}\right\vert \geq 3,$ they remain of the same
order. Thus, we obtain the relation needed.
\end{proof}

\subsection*{Acknowledgements}
 The work of V.V.~was supported by the Grant of
the President of Ukraine for young scientists GP/F8/0045.
 Author thankful to
Professor L.~Pastur for numerous helpful discussions.

\LastPageEnding

\end{document}